# Molecular dynamics study of molecular mobility in catenanes

Ye.V. Tourleigh[1, a], K.V. Shaitan [1, b]

[1]Moscow State University, Faculty of Biology, Dept. of Bioengineering, 119899 Moscow, Russia

[a]yegr@moldyn.ru, [b]shaitan@moldyn.ru

**Abstract.** Molecular machines described in this paper are meant to be such molecular systems that make use of conformational mobility (i.e. hindered rotation around chemical bonds and molecular construction deformations with formation and breakage of nonvalent bonds). Components of molecular machines move mainly by means of restricted diffusion. As an example of molecular machines of a nonbiological nature catenanes (compounds with two interlocked molecular rings) can be proposed. Thus, for example, model catenane ((2)-(cyclo-bis(paraquat-p-phenylene))-(1(2,6)-tetrathiafulvalena-16(1,5)naphtalena-3, 6, 9, 12, 15, 17, 20, 23, 26, 29-decaoxatnacontaphane)-catenane) changes its redox status when an electric field is applied, and rotation of the rings takes place. It occurs with fixation at certain moments of the influence. To find out characteristic properties of rings movements under various external conditions molecular dynamics simulation was carried out. Three cationic forms of the catenane were first subjected to geometrical optimization and quantum chemical calculation.

## Introduction

Drawing macroscopic analogies on the phenomena of molecular level establishes relationships between two approaches to studying structure of substance – large-downward (top-down) and small-upward (bottom-up) ones. In this connection, molecular machines present an image which could be characterized not only by the features that reside in ordinary systems of molecular level, but also by the terms that describe large constructions. Of course, dynamics of a molecular machine, generally, a set of components (molecules, groups of atoms), is stochastic to a great extent [1]. That is caused by movement of elements of the machine by the mechanism of restricted diffusion [2]. Nevertheless, at the appropriate influence components can make movements similar to mechanical – predominance of large-scale displacements in a certain direction and conformational transitions are observed. An experimentally well investigated example of molecular machines is catenanes which are compounds consisting of two interlocked molecular rings [3].

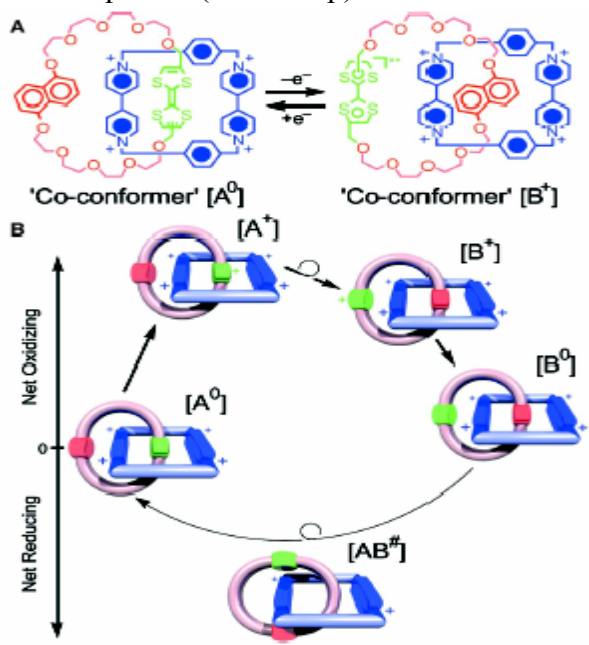

Fig. 1. (*A*) Molecular drawing of the bistable [2]catenane. The voltage-driven circumrotation of co-conformer [A$^0$] to co-conformer [B$^+$] is the basis of the device. (*B*) Proposed mechanochemical mechanism for the operation of the molecular switch. (Reproduced from [4] with courteous permission of the authors).

Practical application of catenanes has been already investigated in light of their possible use as semi-conductor switches in logic circuits of elementary computers [4]. In the proposed model the catenane, being between two electrodes, changes its oxidation status under the action of an external electric field. Herewith, rotation of one ring in the cavity of another occurs accompanied by its fixation at certain moments of the influence (Fig. 1). Such fixation, in principle, is possible owing to catenane being put as a monomolecular film between closely located plane electrodes. As given in Fig. 1, the mechanochemical scheme co-conformer [$A^0$] represents both the ground-state state of the [2]catenane and the "switch open" state of the device (no current flows). When the [2]catenane is oxidized (by applying a bias of -1.5 V ) the tetrathiafulvalene groups (at the right in left ring in [$A^0$]) are ionized and experiences Coulomb repulsion with the tetracationic cyclophane (right ring), resulting in circumrotation of the ring and formation of co-conformer [$B^+$]. When the voltage is reduced to a near-zero bias, co-conformer [$B^0$] is formed, and this represents the "switch closed" state of the device (0.2 µA current flows). Partial reduction of the cyclophane (at an applied bias of +2 V) is necessary to regenerate [$A^0$] co-conformer. Thus, the device can be brought to the closed state (when the current flows) and to the open state (when the current doesn't flow) cyclically.

As a result of the aforesaid, the purpose of the present work is a computer simulation of the dynamics of the model [2]catenane which electroconductive properties are supposed to be connected with its conformation. Investigation methods in silico (computer-simulated) meanwhile represent a unique way to study in detail the dynamic correlations in such systems; and till now molecular dynamics of catenanes hasn't been studied in connection with their functional properties, only semi-empirical quantum chemical calculations [5] and simulation of neutron scattering with use of molecular dynamics [6,7] have been carried out. In the article, the main types of the movements occurring in the catenane are considered.

**Objects and methods of research**

Geometrical optimization of three cationic forms of (2)-(cyclo-bis(paraquat-p-phenylene))-(1(2,6)-tetrathiafulvalena-16(1,5)naphtalena-3,6,9,12,15,17,20,23,26,29-decaoxatriacontaphane)-catenane [8] (cat$A^0$, cat$B^+$ и cat$AB^\#$, Fig. 1) and quantum chemical calculation of the molecular charge distribution of the molecule became the first stage of the work. The restricted Hartree-Fock method with open shells was used, and the optimization was carried in STO-3G basis with descent in TRIM (Trust Image Radius Minimization, [9]), one of the quasi-Newton-Raphson method varieties. The partial charges were calculated according to Loewdin. The whole quantum chemical set of calculations was executed in the software package GAMESS$^©$ (General Atomic and Molecular Electronic System Structure, [10]).

Comparative analysis of existing atom types of the Amber99$^©$ force field used in calculation of classical molecular dynamics was taken further. Taking into account the geometry of the structures optimized to some extent (gradient on the final step of the optimization came to less than 0,001% of the final energy and was of one order of magnitude more than the final gradient recommended for calculation of small molecules), new types of atoms, bonds, valent and torsion angles were assigned. In particular, new types of atoms which form new types of bonds and angles, having intermediate values of the rigidity constant among well-known analogues, were set up in correspondence to all heavy atoms of tetrathiafulvalene and naphthalene units and two bipyridines. Charges were assigned according to the previous quantum chemical calculations.

The following stage of the work was an investigation of the catenane at the level of the molecular mechanical approximation. The calculation was carried out in the package MoDyP [11]. Molecular dynamics was conducted for one chosen molecule of the catenane under different conditions, common of which were the step of integration, 0.001 ps, and combined use of NVT and collision thermostats (average frequency of collisions with virtual particles was 100 per ps per atom, weight of the particles was 18 a.w.u.), and also radiuses of truncating for Coulomb and Van-der-Waals interactions – 15 Å and 12 Å correspondingly. Temperature varied from 300 K to 2000 K. The range of trajectories length was from 10 ns to 1 µs. Also a set of trajectories (each had 10 ns in length), in which the frequency of the external electric field (0.0001 – 10 ns$^{-1}$) was varying, was calculated for cat$A^0$. The choice of the magnitude value of external electric field imposed upon the system in some calculations, 7.2·10$^6$ V/cm, is conditioned on that it corresponds to such field which is created by a point charge 2e- at a distance of 20 Å in vacuum (20 Å is a characteristic size of the catenane; the catenane can be freely placed in a cube with sides of such length). Besides, the dynamics of catA$^0$ with cyclophane ring (CPH) being fixed and with tetrathiafulvalene ring (TTF) being fixed was monitored (lengths of trajectories amount to 100 ns, set temperature was 300 K).

In addition to visual representation of results, the quantitative analysis of the molecules' mobility depending on above-listed parameters was carried out. Time dependences of rotation angles, angular velocities, angular and dipole moments for CPH and TTF, auto- and crosscorrelation functions of these quantities, spectral Fourier expansion on frequency dependences of the quantities were calculated. The angle of rotation for a ring was determined as the angle on which a projection of a radius-vector drawn out of the ring's centre of masses to an atom chosen at one's own rotates in the ring's plane from the initial position. The plane of a ring was set as a bidimensional space stretched on the basis vectors obtained by orthogonalization of two vectors connecting two pairs of atoms in the ring. The chosen vectors are symmetric relative to an axis of the second order symmetry if considered relative to the planar formula of the ring. Four sp3-atoms of carbon diametrically located by pairs in the corners of the "quadrangular" ring were the base for the reference vectors in the CPH ring. In the TTF ring the base was the atoms of oxygen of the crown-ether bridges; these atoms constrain naphthalene and tetrathiafulvalene, the most structurally rigid elements of the ring, and therefore present the most suitable reference points that could be used at assignment of a plane in the ring, structure that is rather flexible as a whole.

**Results and discussion**

Quantum chemical calculations have correctly reflected that catA$^0$ is oxidized to catB$^+$ very due to sulfur of the tetrathiafulvalene unit, as well as at the reduction catA$^0$ → catAB# the nitrogen of bipyridine is reduced first of all. Pyridine rings in catA carry a charge close to +1. Thus, the total charge of CPH is equal to +3.97 and the charge of TTF is equal to +0.03. In much the same manner, in catB$^+$ the charge of CPH is close to +4 and the charge of TTF is close to +1. In catAB$^#$ the charge of CPH ≈ +2, the charge of TTF ≈ 0. π-electron acceptor status of the charged cyclophane ring's bipyridine nitrogens and π-electron donor status of the tetrathiafulvalene ring's nonoxidized sulfur are seen in such a system, as it was noted in [12].

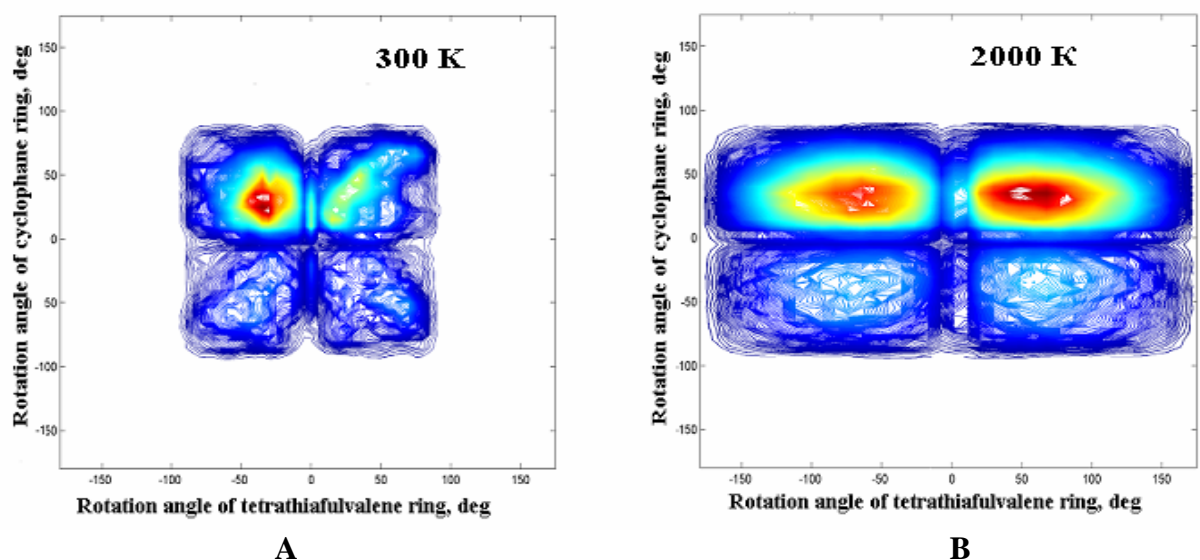

Fig. 2. Density of states of rotation angles of the rings in [A$^0$] co-conformer at 300 K (*A*) and at 2 000 K (*B*). Zero on both axes corresponds to the initial state.

At temperature 300 K rotation of rings on more than 180° occurs in none of the catenane's co-conformers in the period of trajectory length (no more than 1 μs). At 2 000 K such rotation takes place only in catA$^0$ one time per 30 ns on average, and it was observed only for the TTF ring. Taking into account the known experimental data on the energy of activation in similar catenanes (10 kcal/mol), it is possible to assume that at room temperature rotation is possible with times of about several tens of milliseconds. The distribution on angles of rotation of the CPH and TTF rings around their axes is shown in Fig. 2. One can see that with an increase of temperature the area of TTF rotation angle equal to +50° begins to be filled up in addition to the symmetric area with angle -50°. At that, the CPH ring does not change its distribution on angles with increase of temperature. It really points out the importance of TTF's rotation, and not CPH's one, for the cyclic transition according to the scheme on Fig. 1. Consideration of dipole moments of the rings also leads to a similar conclusion. If considering correlation time of dipole moment of a ring as a measure of how greatly electrostatic effects influence the coordination of internal movements in time one can see that with fixed strongly charged CPH the other ring (TTF) moves more coordinately than vice versa (Fig. 3). The fact that the correlation time for CPH's dipole moment almost doesn't depend on the fixing of the rings tells also about initially smaller rotational mobility of CPH.

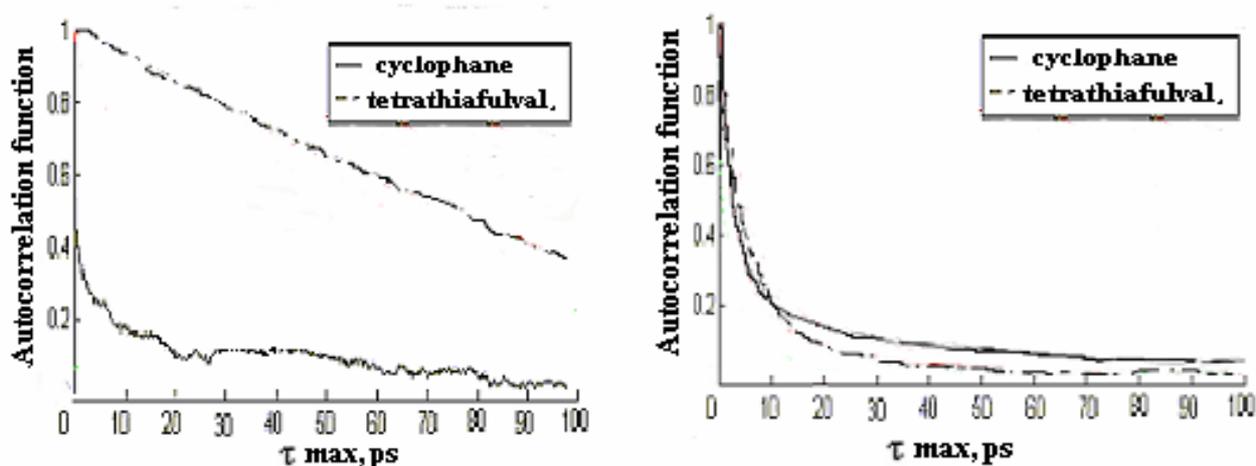

Fig. 3. Comparison of autocorrelation functions of dipole moments for the catA$^0$ system with fixed cyclophane ring (on the left) and fixed tetrathiafulvalene ring (on the right). Temperature 300 K. The dipole moments were calculated relative to the centre of gravity for the whole catenane:

$$\vec{p} = \sum_{i=1}^{n} q_i \cdot (\vec{r}_i - \vec{R}_m)$$

where $q_i$ denotes charge of ith atom in a ring, $\vec{r}_i$ – its radius-vector, $\vec{R}_m$ – radius-vector of the centre of gravity for the whole molecule.

Specified correlation times (tens of picoseconds at standard temperature) of the dipole moments correspond to "trembling" movements of atomic groups in the rings and do not correspond at all to rotary movement of the rings since the former are identical to all cationic forms of the catenane whereas rings in catB$^+$ and catAB$^\#$ during 300 ns haven't made full rotations even at 2000 K, in contrast to the TTF ring in catA$^0$.

The same times can be obtained from dependence of autocorrelation times of dipole moments on the frequency of the external electric field (Fig. 4). The matter is that dipole molecules have time to be oriented under action of the external field if the period of its fluctuations is more than the relaxation time. If the period of the field fluctuations is of the same order of magnitude as the relaxation time molecules have time to return to the initial chaotic state during changing of the field sign. There is an abrupt decrease of orientational polarization in the appropriate area of frequencies. Herewith, energy is partially absorbed by the substance as the field acts against disorienting forces in the substance. This area is an area of abnormal dispersion of electric waves. Absorption of electric waves results in dielectric losses. Abnormal dispersion and dielectric losses arise in liquids at frequencies about $\nu = 1/\tau \approx 10^{11}$ sec$^{-1}$. It follows from the Debye formula:

$$\tau = \frac{4\pi\eta\sigma^3}{kT},$$

where $\eta$ denotes liquid viscosity, and $\sigma$ denotes effective molecular radius. The characteristic quenching time is tens of picoseconds for such molecule as catA$^0$. So decrease of the autocorrelation times of dipole moments (as well as angular ones) occurs at a frequency of about 0.1 ps$^{-1}$ (Fig. 4).

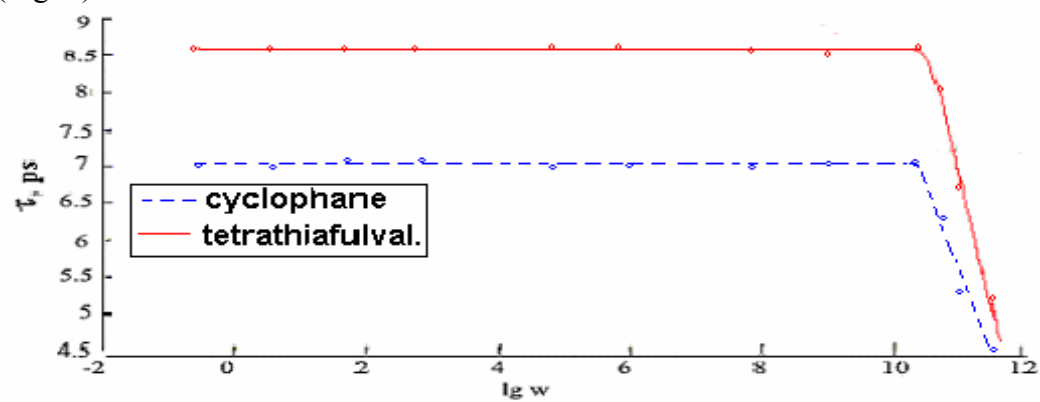

Fig. 4. Dependence of autocorrelation time $\tau$ of dipole moments of both of rings on frequency $w$ of imposed electric field.

Fourier expansion of angular velocities of the rings' rotation (Fig. 5) was used for finding out the movement's characteristic frequencies. It is impossible to pick out definite modes in rotation of the CPH ring; actually, its movement is completely disorderly. The prevalence of low (less than 1 ns$^{-1}$) frequencies over common noise is visible in the rotation of the TTF ring. Nevertheless, the factor determining stochasticity, i.e. common noise, is present. The main maxima lay outside the area of discretizaton for such a Fourier spectrum (less than 1 ms$^{-1}$).

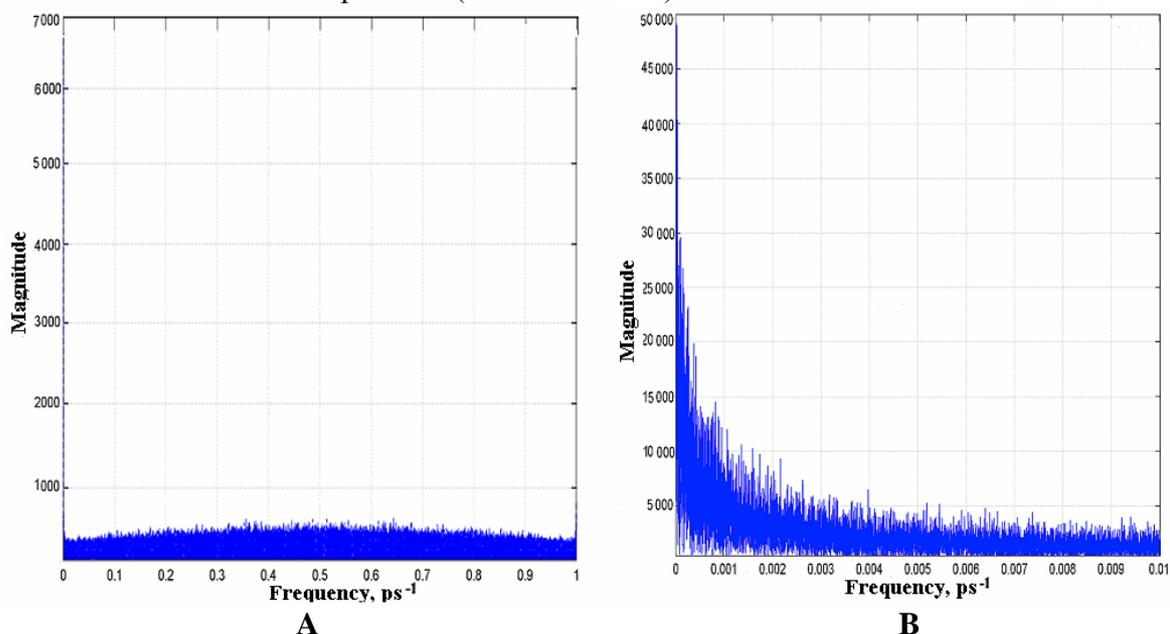

Fig. 5. Fourier expansions of angular rotation velocities of the rings in catA$^0$ at 300 K. (*A*) CPH, (*B*) TTF.

Characteristic values of quenching times for the autocorrelation functions of rotation angles of both of rings in catA$^0$ amount to 900 ps for CPH and 1500 ps for TTF at 300 K (Fig. 6). In this way, large-scale rotations occur in a period much more than the correlation time of rotation angles, and dynamics of rings' movement is determined, substantially, by stochastic rotation of the rings accompanied by bending movements of the crown-ether bridges of TTF.

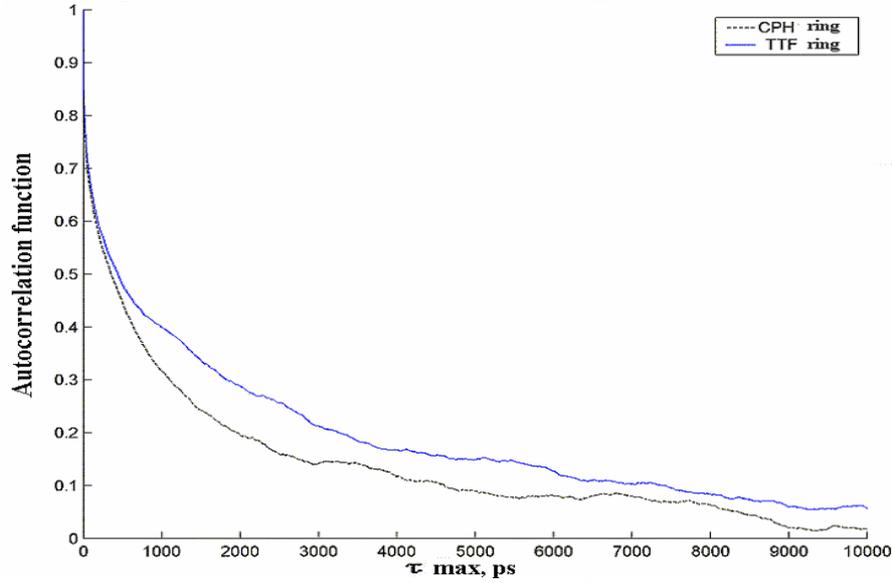

Fig. 6. Autocorrelation functions of rotation angles of the rings in catA$^0$ at 300 K. The functions were taken as follows: $F(\tau) = \langle e^{i\varphi(t)-i\varphi(t+\tau)} \rangle - |\langle e^{i\varphi(t+\tau)} \rangle|^2$ [13], where $\varphi(t)$ denotes rotation angle of the ring. Real parts of the quantities define autocorrelation functions' behavior.

Crosscorrelation of rotation angles is rather weak; maxima are seen on times of nanosecond order (Fig. 7). That is to say there is a coordination at the level of bending movements.

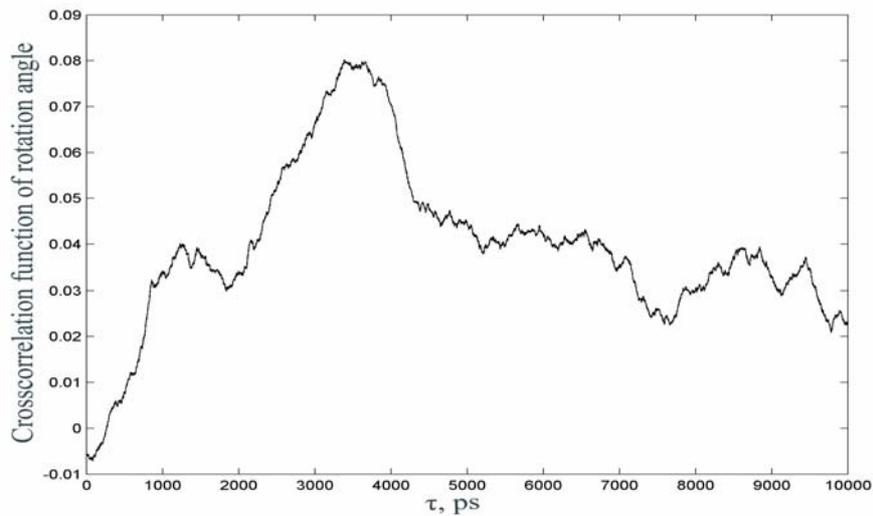

Fig. 7. Crosscorrelation function of rotation angles of the rings in catA$^0$ at 300 K. The function was taken as follows: $F(\tau) = \langle e^{i\varphi(t)-i\varphi(t+\tau)} e^{-i\psi(t)+i\psi(t+\tau)} \rangle - \langle e^{i\varphi(t)-i\varphi(t+\tau)} \rangle \langle e^{-i\psi(t)+i\psi(t+\tau)} \rangle$ [13], where $\varphi(t)$ and $\psi(t)$ denote rotation angles of the rings. Real part of the quantity defines crosscorrelation function's behavior.

**Summary**

The studying of catA$^0$ dynamics shows the presence of two types of constituent rings' movement: large-scale rotation of the TTF ring through angles more than 180º and rapid disorderly rotary movements of both rings. More than one full rotation of the weakly charged TTF ring occurs during 1 ms only at high temperatures, at room temperature it should occur in a period too excessive for machine counting – tens of milliseconds. The other ring does not rotate through significant angles in any investigated cases. In co-conformers catB$^+$ and catAB$^\#$ rings do not rotate greatly during 100 ns even at 2000 K. It remains only to suppose that assigned distribution of charges is an essential factor in the rotation of the rings; but nevertheless, the assignment is well coordinated to known data from electrochemistry [14]. It is necessary to note, that at trial simulation of dynamics of the cationic form catA$^+$ (Fig. 1) the TTF ring makes the CPH ring turn through 180º in the first several nanoseconds, turning thus into co-conformer catB$^+$ (charge distribution was taken identical for both forms). Further reverse rotation of such a scale does not occur during the calculation. Similarly, behavior of catAB$^\#$ is described within the framework of the scheme on Fig. 1 as well – during the calculation (total duration 100 ns) the CPH ring remained mostly in a position between naphthalene and tetratiafulvalene units, i.e. it was strung on the crown-ether bridge.

The movements of different time-scales can be characterized either by statistical functions of rotation angles and angular velocities or by spectral decomposition of the same quantities. Large-scale rotation comes to light the best in Fourier spectra, bending movements occuring on times scales of nanosecond order – with help of correlation analysis. The dielectric relaxation is concerned with "trembling" of the rings and occurs on smaller time scales.

Accordingly, the total rotational movement of the rings in catenane co-conformers catA$^0$ (catB$^0$) and catAB$^\#$ is stochastic but for the TTF ring it includes selected degrees of freedom with the prevalence of large-scale rotational movements in frequency contribution to the total spectrum. For the most charged forms, catA$^+$ and catB$^+$, rotation through angles more than 180º is absent or occurs once at the transition catA$^+$ → catB$^+$. Probably, it is concerned with the strong influence of redox-status.

Partial financial support from RF Ministry of Education (grant И0431, program «Integratsia», grant № 01.106.11.0001) and RFBR (grant № 04-04-49645) is gratefully acknowledged.